\documentclass[preprintnumbers,amsmath,amssymb,showpacs]{revtex4}
\usepackage{graphicx}% Include figure files
\usepackage{dcolumn}% Align table columns on decimal point
\usepackage{bm}% bold math
\usepackage{booktabs}
\usepackage{CJK}
\usepackage{subfigure}
\usepackage{qcircuit}
\usepackage{color}
\usepackage{braket}
\usepackage{listings}
\usepackage{soul}
\usepackage{xcolor}
\soulregister\cite7 % Õë¶Ô\cite ÃüÁî
\soulregister\ref7 % Õë¶Ô\ref ÃüÁî
\usepackage{color, soul} %ÓÃcolor, ºÍ soul °ü
\setstcolor{yellow}

\begin{document}

\title{Symmetric quantum joint measurements on multiple qubits}

%\author{***}
\author{Dong Ding$^{1}$}
%\email{dingdong@ncist.edu.cn}
\author{Ying-Qiu He$^{1}$}
\email{heyq@ncist.edu.cn}
\author{Ting Gao$^{2}$}
\email{gaoting@hebtu.edu.cn}
\author{Feng-Li Yan$^{3}$}
\email{flyan@hebtu.edu.cn}

\affiliation {
$^1$ College of Science, North China Institute of Science and Technology, Beijing 101601, China\\
$^2$ School of Mathematical Sciences, Hebei Normal University, Shijiazhuang 050024, China\\
$^3$ College of Physics, Hebei Key Laboratory of Photophysics Research and Application,  Hebei Normal University, Shijiazhuang 050024, China
}
\date{\today}

\begin{abstract}
We investigate the generalization of symmetric quantum joint measurements on multiple qubits.
We first describe a method for constructing a symmetric joint measurement basis for three qubits by utilizing single-qubit states corresponding to the four vertices of a tetrahedron on the Bloch sphere. We demonstrate the expected tetrahedral symmetry of the current measurement basis and discuss its application in a trilocal star-shaped network.
This architecture enables us to generalize the two-qubit symmetric joint measurement to an
$n$-qubit version, preserving the tetrahedral or hexahedral symmetry.
\end{abstract}


%Keywords: quantum measurement; symmetric join measurement; quantum circuit

%03.65.Ud:Entanglement and quantum nonlocality; 03.67.-a:Quantum information;
%42.50.-p:Quantum optics; 03.67.Lx Quantum computation architectures and implementations
%03.67.Hk Quantum communication; 03.67.Mn Entanglement measures, witnesses, and other characterizations
%
\maketitle

\section{Introduction}

Quantum measurement is a cornerstone of quantum mechanics and underlies the concepts of quantum computation and quantum information processing \cite{NC2000,Quantum-entanglement}.
It includes both single-system measurements and composite-system joint measurements. The well-known Bell state measurement (BSM) and its generalized forms serve as the canonical method for detecting multi-qubit systems \cite{Entanglement-detection2009}.

In 2019, a novel symmetric joint measurement for two qubits, termed the {\it elegant joint measurement} (EJM), was proposed by Gisin \cite{Gisin-EJM2019}.
The symmetry of the EJM basis is mainly reflected in that both the single-qubit ingredient and the reduced states are precisely associated with the four vertices of a specified tetrahedron.
After Gisin's idea, Tavakoli \emph{et al.} \cite{TGB-EJM2021} proposed the parameterized EJM by introducing a real parameter $\theta \in [0,\pi/2]$. This formulation retains the elegant symmetry, allowing it to encompass both the EJM and the conventional BSM.
The EJM basis states were initially designed for testing non-bilocal correlations  \cite{Phys.Rev.Lett.104:170401(2010),PhysRevA.90.062109(2014),PhysRevLett.120.140402(2018),Phys.Rev.Lett.123:140503(2019), FNN-PRL2022, Tavakoli-network2022} and it has been experimentally confirmed via superconducting quantum processors \cite{BGT2021IBM} as well as hyperentangled photons \cite{GuoPRL-EJM2022}.
Recently, Del Santo \emph{et al.} \cite{PRR-EJM2024} have proposed an in-depth study on two-qubit entangled measurements. In their work, the authors offered a complete classification of two-qubit bases that have the same degree of entanglement. They also suggested that an iso-entangled basis for two qubits typically depends on three parameters.
In view of this, He \emph{et al.} \cite{3-parameter-EJM2025} extended the single-parameter EJM to three-parameter EJM. It offers a broader range of points on the Bloch sphere (unit vectors) that are suitable for constructing the generalized EJM basis.
However, compared to the generalized BSM (or GHZ-state measurement), it does not appear to be an
issue that can be naturally generalized, due to the constraints imposed by tetrahedral symmetry.

In the present paper, we focus on multi-qubit symmetric joint measurements. We first define a three-qubit EJM basis by using the three-parameter EJM and verify their tetrahedral symmetry.
Next, we apply the three-qubit EJM to the trilocal star-shaped network, effectively demonstrating network nonlocality.
Finally, we provide a method to develop the multi-qubit EJM bases and analyze its symmetry.

\section{Symmetric joint measurements on two qubits}

We here review two-qubit symmetric joint measurements.
Consider four pure single-qubit states
\begin{eqnarray}
|m^{}_{i}\rangle=\frac{1}{\sqrt{2}}   (\sqrt{1 + z_{i}}\text{e}^{-\text{i}\varphi_{i}/2}|0\rangle
+ \sqrt{1 - z_{i}}\text{e}^{\text{i}\varphi_{i}/2}|1\rangle), ~ i=0,1,2,3,
\end{eqnarray}
each of which is respectively associated with a unit vector
\begin{eqnarray} \label{vec-m}
\vec{m}_{i} = \langle m^{}_{i}|\vec{\sigma}|m^{}_{i}\rangle = (\sqrt{1-z_{i}^{2}}\cos \varphi_{i},\sqrt{1-z_{i}^{2}}\sin \varphi_{i},z_{i}).
\end{eqnarray}
Here, $\vec{\sigma}=(\sigma_{x},\sigma_{y},\sigma_{z})$ is the vector of Pauli matrices, the real parameters $|z_{i}| \leq 1$ and $\varphi_{i} \in [-\pi,\pi]$.

Choosing $\varphi_{0}=\pi/4$, $\varphi_{1}=-\pi/4$, $\varphi_{2}=3\pi/4$, $\varphi_{3}=-3\pi/4$, $z_{0}=z_{3}=1/\sqrt{3}$, $z_{1}=z_{2}=-1/\sqrt{3}$, i.e.,
\begin{eqnarray} \label{m-1234}
\vec{m}_{0}= (1,1,1)/\sqrt{3},   ~~  \vec{m}_{1}= (1,-1,-1)/\sqrt{3}, ~~
\vec{m}_{2}= (-1,1,-1)/\sqrt{3}, ~~  \vec{m}_{3}= (-1,-1,1)/\sqrt{3},
\end{eqnarray}
one can define a set of two-qubit basis, parameter-free EJM \cite{Gisin-EJM2019}, as
\begin{eqnarray} \label{Gisin-EJM-basis}
|\Phi_{i}\rangle=\frac{1}{2\sqrt{2}}
[(\sqrt{3}+1)|m_{i},-m_{i}\rangle+(\sqrt{3}-1)|-m_{i},m_{i}\rangle], ~~i=0,1,2,3,
\end{eqnarray}
where $|-m_{i}\rangle = ({1}/\sqrt{2}) (\sqrt{1-z_{i}}\text{e}^{-\text{i}\varphi_{i}/2}|0\rangle
- \sqrt{1 + z_{i}}\text{e}^{\text{i}\varphi_{i}/2}|1\rangle)$ is orthogonal to $|m_{i}\rangle$.

It can be further parameterized by incorporating a variable $\theta \in [0,\pi/2]$, single-parameter EJM \cite{TGB-EJM2021}, as
\begin{eqnarray} \label{TGB-EJM-basis}
|\Phi_{i}^{\theta}\rangle=\frac{1}{2\sqrt{2}}  [(\sqrt{3}+\text{e}^{\text{i}\theta_{}})|m_{i},-m_{i}\rangle
+(\sqrt{3}-\text{e}^{\text{i}\theta_{}})|-m_{i},m_{i}\rangle],
\end{eqnarray}
such that it seamlessly interpolates between the EJM ($\theta = 0$) and the normal BSM ($\theta = \pi/2$).

Note that a two-qubit symmetric joint measurement is generally characterized by three parameters \cite{PRR-EJM2024}.
Consider three real parameters $1/\sqrt{3} \leq |z|\leq 1$, $\varphi \in [-\pi, \pi]$ and $\theta_{} \in [0, \pi/2]$.
A more generalized symmetric joint measurement, three-parameter EJM \cite{3-parameter-EJM2025}, is then given by
\begin{eqnarray} \label{Phi0-1-z}
|\Phi_{0}(z,\varphi,\theta)\rangle &=& \frac{1-\text{i}\sqrt{3z^{2}-1}}{2\sqrt{3z^{2}}}
[\text{e}^{-\text{i}(\varphi-\varphi_{z})}|00\rangle
- r_{+}^{\theta} |01\rangle - r_{-}^{\theta} |10\rangle
- \text{e}^{\text{i}(\varphi-\varphi_{z})}|11\rangle],
\end{eqnarray}
\begin{eqnarray} \label{Phi1-1-z}
|\Phi_{1}(z,\varphi,\theta)\rangle &=& \frac{1-\text{i}\sqrt{3z^{2}-1}}{2\sqrt{3z^{2}}}
[-\text{i}\text{e}^{-\text{i}(\varphi-\varphi_{z})}|00\rangle
+ r_{-}^{\theta} |01\rangle + r_{+}^{\theta} |10\rangle  -\text{i}\text{e}^{\text{i}(\varphi-\varphi_{z})}|11\rangle],
\end{eqnarray}
\begin{eqnarray} \label{Phi2-1-z}
|\Phi_{2}(z,\varphi,\theta)\rangle &=& \frac{1-\text{i}\sqrt{3z^{2}-1}}{2\sqrt{3z^{2}}}
[-\text{e}^{-\text{i}(\varphi-\varphi_{z})}|00\rangle
- r_{+}^{\theta} |01\rangle - r_{-}^{\theta} |10\rangle +\text{e}^{\text{i}(\varphi-\varphi_{z})}|11\rangle],
\end{eqnarray}
\begin{eqnarray} \label{Phi3-1-z}
|\Phi_{3}(z,\varphi,\theta)\rangle &=& \frac{1-\text{i}\sqrt{3z^{2}-1}}{2\sqrt{3z^{2}}}
[\text{i}\text{e}^{-\text{i}(\varphi-\varphi_{z})}|00\rangle
+ r_{-}^{\theta} |01\rangle + r_{+}^{\theta} |10\rangle
+ \text{i}\text{e}^{\text{i}(\varphi-\varphi_{z})}|11\rangle],
\end{eqnarray}
where $\varphi_{z}=\text{arg}[(\sqrt{1-z^{2}}+\text{i}\sqrt{3z^{2}-1})/(\sqrt{2}|z|)]$ and $r_{\pm }^{\theta} =(1\pm \text{e}^{\text{i}\theta})/\sqrt{2}$.
It is not difficult to see that, if one takes $\varphi_{}-\varphi_{z}=\pi/4$, then the three-parameter EJM can naturally be reduced to single-parameter EJM.

From the perspective of quantifying entanglement, all the EJM basis states have the same degree of entanglement. Furthermore, by computing the partial traces one can observe that these EJM bases have elegant symmetry properties:
(i) The single-qubit reductions, obtained by tracing out one of the qubits, are mirror images of each other; and (ii) the four points, to which the reduced states of the EJM basis point, form a tetrahedron inside the Bloch sphere.

\section{Symmetric joint measurements on three qubits}

\subsection{Definition of three-qubit measurement basis}

Consider four real parameters $1/\sqrt{3} \leq |z|\leq 1$, $\varphi \in [-\pi, \pi]$, $\theta_{} \in [0, \pi/2]$, and $\gamma \in [0, \pi/2]$, and let $\varphi_{0}=\varphi_{}$, $\varphi_{1}=\varphi_{}+\pi/2$, $\varphi_{2}=\varphi_{}+\pi$, $\varphi_{3}=\varphi_{}-\pi/2$, $z_{0}=z_{2}=z$, $z_{1}=z_{3}=-z$.
Define eight three-qubit pure states
\begin{eqnarray} \label{Psi-i-0}
|\Psi_{i}^{0}\rangle &=&
  \cos \gamma |\Phi_{i}\rangle|m_{i}\rangle
+ (-1)^{[\frac{i}{2}]} \sin \gamma |\Phi'_{i}\rangle|-m_{i}\rangle, ~  i=0,1,2,3
\end{eqnarray}
and
\begin{eqnarray} \label{Psi-i-1}
|\Psi_{i}^{1}\rangle &=&
  \cos \gamma |\Phi_{i}\rangle|-m_{i}\rangle
- (-1)^{[\frac{i}{2}]} \sin \gamma |\Phi'_{i}\rangle|m_{i}\rangle, ~  i=0,1,2,3,
\end{eqnarray}
where
\begin{eqnarray} \label{3parameter-EJM-basis}
|\Phi_{i}\rangle &=& \frac{1-\text{i}\sqrt{3z^{2}-1}}{2\sqrt{3z^{2}}}
[\text{e}^{-\text{i}(\varphi_{i}-\varphi_{z})}|00\rangle
- \frac{1}{\sqrt{2}}[(-1)^{i} + \text{e}^{\text{i}\theta}] |01\rangle
- \frac{1}{\sqrt{2}}[(-1)^{i} - \text{e}^{\text{i}\theta}] |10\rangle -\text{e}^{\text{i}(\varphi_{i}-\varphi_{z})}|11\rangle]
\end{eqnarray}
and
\begin{eqnarray} \label{3parameter-EJM-basis'}
|\Phi'_{i}\rangle &=& \frac{1-\text{i}\sqrt{3z^{2}-1}}{2\sqrt{3z^{2}}}
[-\text{e}^{-\text{i}(\varphi_{i}-\varphi_{z})}|00\rangle
- \frac{1}{\sqrt{2}}[(-1)^{i} + \text{e}^{\text{i}\theta}] |01\rangle
- \frac{1}{\sqrt{2}}[(-1)^{i} - \text{e}^{\text{i}\theta}] |10\rangle +\text{e}^{\text{i}(\varphi_{i}-\varphi_{z})}|11\rangle]
\end{eqnarray}
are all the three-parameter two-qubit EJM basis states satisfying  $\langle \Phi_{i}|\Phi_{j}\rangle = \delta_{ij}$ and $\langle \Phi'_{i}|\Phi'_{j}\rangle = \delta_{ij}$, $i, j=0,1,2,3$, $\varphi_{z}=\text{arg}[(\sqrt{1-z^{2}}+\text{i}\sqrt{3z^{2}-1})/(\sqrt{2}|z|)]$, $|\pm m_{i}\rangle =(\sqrt{1 \pm z_{i}}\text{e}^{-\text{i}\varphi_{i}/2}|0\rangle \pm \sqrt{1 \mp z_{i}}\text{e}^{\text{i}\varphi_{i}/2}|1\rangle)/\sqrt{2}$ are single-qubit basis states, as described earlier.
Here, the floor function $[\frac{i}{2}]=0$ for $i=0,1$, and $[\frac{i}{2}]=1$ for $i=2,3$.
Then, $\{|\Psi_{i}^{k}\rangle\}$ form a three-qubit EJM basis, where $i=0,1,2,3$, and $k=0,1$.

The proof of the orthogonality and completeness is provided in Appendix A. We next see the tetrahedral symmetry of the present three-qubit basis states.

\subsection{Entanglement measure and tetrahedral symmetry}

An important three-qubit measure of entanglement for pure states is the three-tangle introduced in Ref. \cite{three-tangle-2000}. We calculate the three-tangle of the state $|\Psi_{i}^{k}\rangle$ by
\begin{eqnarray} \label{3-tangle-expression}
\tau(|\Psi_{i}^{k}\rangle) &=&
4|(a_{000}a_{111}-a_{001}a_{110})^{2}+(a_{010}a_{101}-a_{100}a_{011})^{2}
+4(a_{000}a_{110}a_{101}a_{011}+a_{111}a_{001}a_{010}a_{100})
\nonumber \\ &&
-2(a_{000}a_{111}+a_{001}a_{110})(a_{010}a_{101}+a_{100}a_{011})|,
\end{eqnarray}
where the coefficients $a_{j_{1}j_{2}j_{3}}=\langle j_{1}j_{2}j_{3}|\Psi_{i}^{k}\rangle, ~ j_{1},j_{2},j_{3}=0,1$.
A straightforward calculation shows that
\begin{eqnarray} \label{3-tangle}
\tau(|\Psi_{i}^{k}\rangle) &=& \sin^{2}(2\gamma) \sin\theta, ~~ i=0,1,2,3, ~k=0,1.
\end{eqnarray}

Based on this observation, the current EJM basis is a three-qubit iso-entangled basis \cite{PRA108-022220}. These basis states exhibit the same degree of entanglement parameterized by $\gamma$ and $\theta$, ranging continuously from 0 to 1.
Typically, if one takes $\gamma=\pi/4$ and $\theta=\pi/2$, then it provides the maximally entangled measurement with $\tau=1$; while for $\gamma=0$  (or $\pi/2$) or $\theta=0$, we have $\tau=0$ corresponding to the product states.

We now reveal its inherent symmetry by calculating the partial traces (single-qubit reductions of these basis states).
By using the results
$\langle m_{i}|\vec{\sigma}| m_{i}\rangle = - \langle -m_{i}|\vec{\sigma}| -m_{i}\rangle = \vec{m}_{i}$ and
$\langle \Phi_{i}|\vec{\sigma} \otimes I |\Phi_{i}\rangle = -\langle \Phi_{i}| I \otimes \vec{\sigma} |\Phi_{i}\rangle
= (\cos \theta / \sqrt{2})
(\cos (\varphi_{i}-\varphi_{z}), \sin (\varphi_{i}-\varphi_{z}), (-1)^{i}/\sqrt{2})$, $i=0,1,2,3$,
we have
\begin{eqnarray} \label{s-I-I}
\langle \Psi_{i}^{k}|\vec{\sigma} \otimes I \otimes I |\Psi_{i}^{k}\rangle =
\frac{1}{2}[\sqrt{1+2\cos^{2} (2\gamma)}\cos \theta]  \vec{m'}_{i},
\end{eqnarray}
\begin{eqnarray} \label{I-s-I}
\langle \Psi_{i}^{k}|I \otimes \vec{\sigma} \otimes I|\Psi_{i}^{k}\rangle =
- \frac{1}{2}[\sqrt{1+2\cos^{2} (2\gamma)}\cos \theta] \vec{m'}_{i},
\end{eqnarray}
and
\begin{eqnarray} \label{I-I-s}
\langle \Psi_{i}^{k}|I \otimes I\otimes \vec{\sigma}|\Psi_{i}^{k}\rangle
= (-1)^{k}\cos (2\gamma) \vec{m}_{i},
\end{eqnarray}
where $\vec{m}_{i}$ (see Eq. (\ref{vec-m}))
and
\begin{eqnarray} \label{}
\vec{m'}_{i}=\frac{1}{\sqrt{1+2\cos^{2} (2\gamma)}}
(\sqrt{2}\cos (2\gamma)\cos(\varphi_{i}-\varphi_{z}),
\sqrt{2}\cos (2\gamma)\sin(\varphi_{i}-\varphi_{z}),(-1)^{i})
\end{eqnarray}
are two unit vectors that can respectively configure a tetrahedron on the Bloch sphere.
A detailed derivation of the Eqs. (\ref{s-I-I})--(\ref{I-I-s}) is presented in Appendix B.

The symmetry of the current basis states is reflected in the following two aspects:
(i) The reductions of the EJM basis states result in two pairs of mirror-image tetrahedra: one pair, with a radius (circumradius) of $\cos(2\gamma)$, has its four vertices aligned with the vectors $\vec{m}_{i}$ and $-\vec{m}_{i}$; the other pair, with a radius of $(1/2)\sqrt{1+2\cos^{2} (2\gamma)}\cos \theta$, has its vertices directed towards the vectors $\vec{m'}_{i}$ and $-\vec{m'}_{i}$.
(ii) The cumulative sum of all vectors associated with the reduced states amounts to zero, i.e.,
\begin{eqnarray} \label{}
\sum_{i,k}(\langle \Psi_{i}^{k}|\vec{\sigma} \otimes I \otimes I |\Psi_{i}^{k}\rangle
+ \langle \Psi_{i}^{k}|I \otimes \vec{\sigma} \otimes I|\Psi_{i}^{k}\rangle
+ \langle \Psi_{i}^{k}|I \otimes I\otimes \vec{\sigma}|\Psi_{i}^{k}\rangle) = 0.
\end{eqnarray}

It is not difficult to see that, the parameters $\gamma$, $z$, and $\varphi_{}$ determine the positions to which the vectors point.
To be more specific, if one takes $\gamma=0$ and $z=1/\sqrt{3}$, then the unit vectors $\vec{m'}_{i}=\vec{m}_{i}$, and thus all reductions of the EJM basis states direct towards one of the same mirror-image vectors
$\pm(\sqrt{2}\cos \varphi_{i}, \sqrt{2}\sin \varphi_{i}, (-1)^{i})/\sqrt{3}$; if one takes $\gamma=\pi/4$, then a pair of tetrahedra reduces to a point with a radius of zero, and the other pair of tetrahedra reduces to two mirror-image points $(0,0,1)$ and $(0,0,-1)$.
While the parameters $\theta$ and $\gamma$ affect the modulus of these vectors,
in particular, if one takes $(1/2)\sqrt{1+2\cos^{2} (2\gamma)}\cos \theta = \cos(2\gamma)$, then the radii of these tetrahedra are equal. For $\theta=\pi/2$ and $\gamma=\pi/4$, related to the maximally entangled measurement $\tau=1$, all radii of the tetrahedra are zero; for $\theta=0$ (thus $\tau=0$) and $\gamma=\pi/8$, then all the tetrahedra have equal radii of $1/\sqrt{2}$.

\section{An application to a trilocal star network}

\subsection {The star-network configuration and trilocal inequality}

\begin{figure}[h]
      \centering
      \includegraphics[width=2.6in]{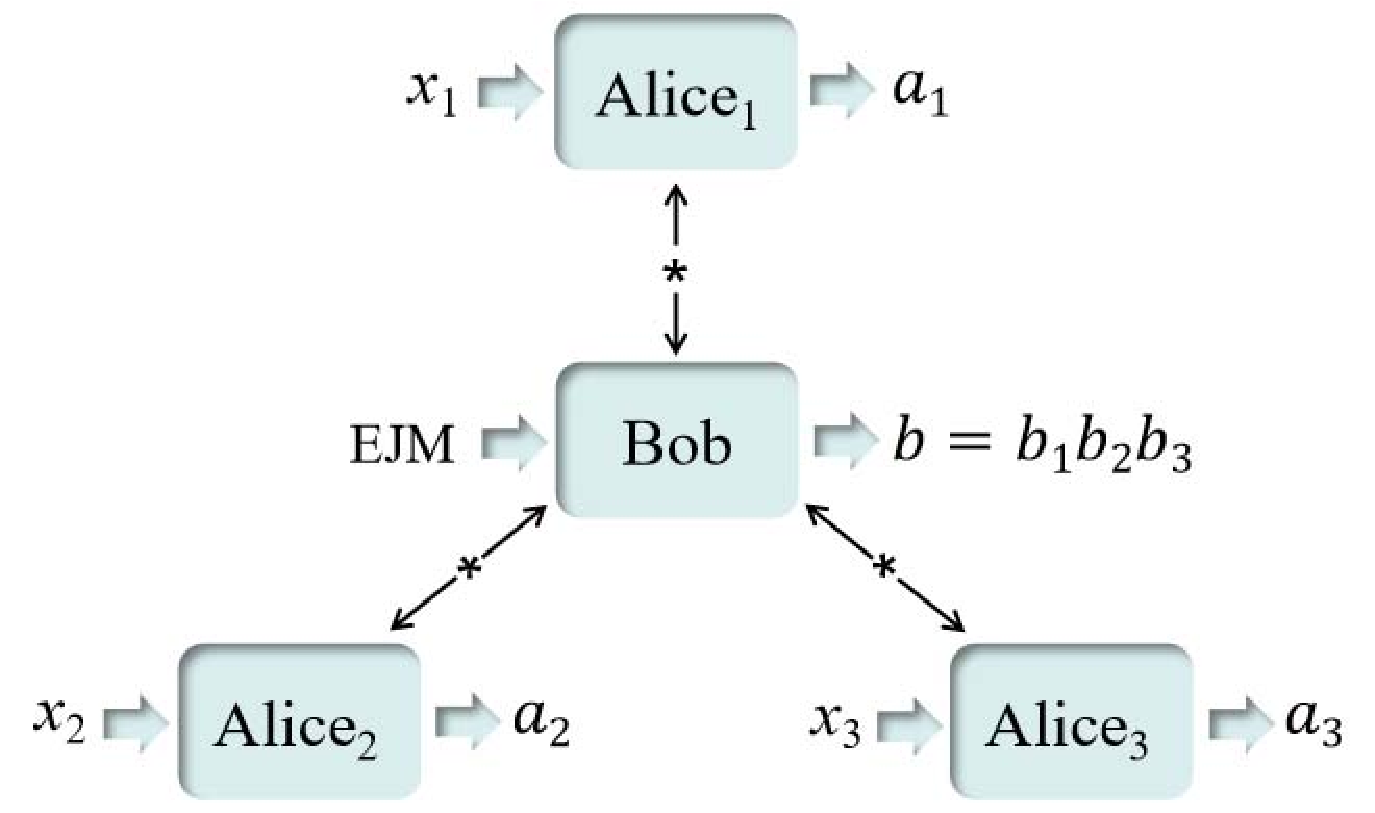}
      \caption{A trilocal star-shaped network with three extremal parties (Alice$_{1}$, Alice$_{2}$, and Alice$_{3}$) preforming dichotomic measurements, respectively, and the central party Bob carrying out a fixed three-qubit joint measurement.}
      \label{star.eps}
\end{figure}

In this section, we investigate the star-network nonlocalities using the current EJM basis.
Consider a trilocal star-shaped network with three extremal parties Alices ($A_{i}, i=1,2,3$) and a central party Bob ($B$), as shown in Fig.\ref{star.eps}.
Suppose that there are three independent (space-like separated) two-particle sources, each of which sends one particle to $A_{i}$ and $B$, respectively.
Each extremal party $A_{i}$ performs a dichotomic measurement with inputs $x_{i}$ and outputs $a_{i}$ (where $x_{i}, a_{i}=0,1$), while the central party $B$ always performs a fixed three-qubit joint measurement and then obtains one of eight possible outputs $b_{}=b_{1}b_{2}b_{3}$ where $b_{1},b_{2},b_{3}=0,1$.

In such a case, consider the trilocal correlation quantities \cite{PhysRevA.90.062109(2014)}
\begin{eqnarray}\label{I-m}
I_{m}=\frac{1}{2^{3}}\sum_{x_{1},x_{2},x_{3}} (-1)^{g_{m}(x_{1},x_{2},x_{3})}\langle A_{x_{1}}^{1}A_{x_{2}}^{2}A_{x_{3}}^{3}B^{m}\rangle, ~ m = 1,2,3,4,
\end{eqnarray}
with the correlator
\begin{eqnarray}\label{correlator}
\langle A_{x_{1}}^{1}A_{x_{2}}^{2}A_{x_{3}}^{3}B^{m} \rangle = \sum_{a_{1},a_{2},a_{3},b^{1},b^{2},b^{3},b^{4}} (-1)^{a_{1}+a_{2}+a_{3}+b^{m}}  P(a_{1}a_{2}a_{3}b^{1}b^{2}b^{3}b^{4}|x_{1}x_{2}x_{3}),
\end{eqnarray}
where
$g_{1}(x_{1},x_{2},x_{3})=0$, $g_{2}(x_{1},x_{2},x_{3})=x_{1}+x_{2}$, $g_{3}(x_{1},x_{2},x_{3})=x_{1}+x_{3}$, $g_{4}(x_{1},x_{2},x_{3})=x_{2}+x_{3}$,
each $b^{m}$ (with $b^{m}=0,1$), corresponding to the operator $B^{m}$, is a processed bit derived from the raw outputs $b$,
and $P(a_{1}a_{2}a_{3}b^{1}b^{2}b^{3}b^{4}|x_{1}x_{2}x_{3})$ is a probability distribution on the trilocal star network.

Then, in the current trilocality configuration, the following inequality
\begin{eqnarray}\label{trilocal-inequality}
\sum_{m=1}^{4} |I_{m}|^{1/3} \leq 2
\end{eqnarray}
holds for every trilocal probability distributions \cite{PhysRevA.90.062109(2014)}.
If a quantum prediction exceeds the classical constraint, i.e., the inequality is violated by a quantum system, then the star-network exhibits quantum nonlocality.

\subsection {Quantum violations}

Suppose that each quantum source produces two-qubit pure state
$|\Psi^{+}\rangle = (|01\rangle + |10\rangle)/\sqrt{2}$.
For each of the three extremal parties Alices, choose $A^{i}_{0}=(\sigma_{x}+\sigma_{z})/\sqrt{2}$ or $A^{i}_{1}=(\sigma_{x}-\sigma_{z})/\sqrt{2}$ on their single-qubit subsystem.
The central party Bob performs a fixed EJM on her three-qubit subsystem.
In detail, let
\begin{eqnarray} \label{B-m}
B^{m}=M^{m}_{0}-M^{m}_{1}, ~ m=1,2,3,4,
\end{eqnarray}
where
\begin{eqnarray} \label{M-m-b}
M^{m}_{b}= \sum_{b_{1}b_{2}b_{3}} \delta_{b,b^{m}} |\Psi_{b_{1}b_{2}b_{3}}\rangle \langle \Psi_{b_{1}b_{2}b_{3}}|, ~ m=1,2,3,4, ~ b=0,1, ~ b_{1}, b_{2},b_{3}=0,1,
\end{eqnarray}
and bit $b^{m}=b^{m}(b_{1}, b_{2},b_{3})$. To proceed, we define the bits
\begin{eqnarray} \label{b123}
b^{1}=b_{2} \oplus b_{3} \oplus 1, ~     b^{2}= b_{3}, ~
b^{3}=b_{1} \oplus b_{3} \oplus 1, ~     b^{4}=b_{1} \oplus b_{2} \oplus b_{3} \oplus 1,
\end{eqnarray}
and set
\begin{eqnarray} \label{Psi-000-00}
&&
|\Psi_{000}\rangle=|\Psi_{0}^{0}\rangle, ~ |\Psi_{001}\rangle=|\Psi_{0}^{1}\rangle, ~
|\Psi_{010}\rangle=|\Psi_{1}^{0}\rangle, ~ |\Psi_{011}\rangle=|\Psi_{1}^{1}\rangle, \\ \nonumber &&
|\Psi_{100}\rangle=|\Psi_{2}^{0}\rangle, ~ |\Psi_{101}\rangle=|\Psi_{2}^{1}\rangle,~
|\Psi_{110}\rangle=|\Psi_{3}^{0}\rangle, ~ |\Psi_{111}\rangle=|\Psi_{3}^{1}\rangle.
\end{eqnarray}

In fact, such a joint measurement allows Bob to perform entanglement swapping to three distant Alices.
To see this, we can rewrite the combined system including six qubits in the EJM basis, as
\begin{eqnarray} \label{}
  |\Psi^{+}\rangle_{A_{1}B_{1}}|\Psi^{+}\rangle_{A_{2}B_{2}}|\Psi^{+}\rangle_{A_{3}B_{3}}
= \frac{1}{2\sqrt{2}} \sum_{b_{1}b_{2}b_{3}} |\widetilde{\Psi}_{b_{1}b_{2}b_{3}}\rangle_{A_{1}A_{2}A_{3}} |\Psi_{b_{1}b_{2}b_{3}}\rangle_{B_{1}B_{2}B_{3}},
\end{eqnarray}
where $|\widetilde{\Psi}_{b_{1}b_{2}b_{3}}\rangle = \sigma_{x}\sigma_{x}\sigma_{x} |\Psi_{b_{1}b_{2}b_{3}}\rangle^{\ast}$,
whose components are complex conjugates of the corresponding components of the state $|\Psi_{b_{1}b_{2}b_{3}}\rangle$, followed by applying a set of NOT gates, one for each of the three qubits.

In this situation, the correlator reads
\begin{eqnarray}\label{}
\langle A_{x_{1}}^{1}A_{x_{2}}^{2}A_{x_{3}}^{3}B^{m} \rangle = \frac{1}{2^{3}}  {\rm tr}[\sum_{b_{1}b_{2}b_{3},b'_{1}b'_{2}b'_{3}} (|\widetilde{\Psi}_{b_{1}b_{2}b_{3}}\rangle \langle \widetilde{\Psi}_{b'_{1}b'_{2}b'_{3}}| \otimes |\Psi_{b_{1}b_{2}b_{3}}\rangle \langle \Psi_{b'_{1}b'_{2}b'_{3}}|)(A_{x_{1}}^{1}\otimes A_{x_{2}}^{2}\otimes A_{x_{3}}^{3} \otimes B^{m})].
\end{eqnarray}
Noting that
$A^{i}_{0}+A^{i}_{1} = \sqrt{2} \sigma_{x}$ and $A^{i}_{0}-A^{i}_{1} = \sqrt{2} \sigma_{z}$,
then we have
\begin{eqnarray}\label{}
I_{m} &=& \frac{1}{16\sqrt{2}}  {\rm tr}[\sum_{b_{1}b_{2}b_{3},b'_{1}b'_{2}b'_{3}} (|\widetilde{\Psi}_{b_{1}b_{2}b_{3}}\rangle \langle \widetilde{\Psi}_{b'_{1}b'_{2}b'_{3}}| \otimes |\Psi_{b_{1}b_{2}b_{3}}\rangle \langle \Psi_{b'_{1}b'_{2}b'_{3}}|)(A^{m} \otimes B^{m})],
\end{eqnarray}
where $A^{1} = \sigma_{x}\sigma_{x}\sigma_{x}$, $A^{2} = \sigma_{z}\sigma_{z}\sigma_{x}$,
$A^{3} = \sigma_{z}\sigma_{x}\sigma_{z}$, and $A^{4} = \sigma_{x}\sigma_{z}\sigma_{z}$.
Substituting Eqs. (\ref{B-m}) and (\ref{M-m-b}) into this expression yields
\begin{eqnarray}\label{I-m}
I_{m} = \frac{1}{16\sqrt{2}} (\sum_{b_{1}b_{2}b_{3}} \delta_{0,b^{m}} \langle \widetilde{\Psi}_{b_{1}b_{2}b_{3}}|A^{m}|\widetilde{\Psi}_{b_{1}b_{2}b_{3}} \rangle - \sum_{b_{1}b_{2}b_{3}} \delta_{1,b^{m}} \langle \widetilde{\Psi}_{b_{1}b_{2}b_{3}}|A^{m}|\widetilde{\Psi}_{b_{1}b_{2}b_{3}} \rangle).
\end{eqnarray}

\begin{figure}[h]
      \centering
      \includegraphics[width=4.5in]{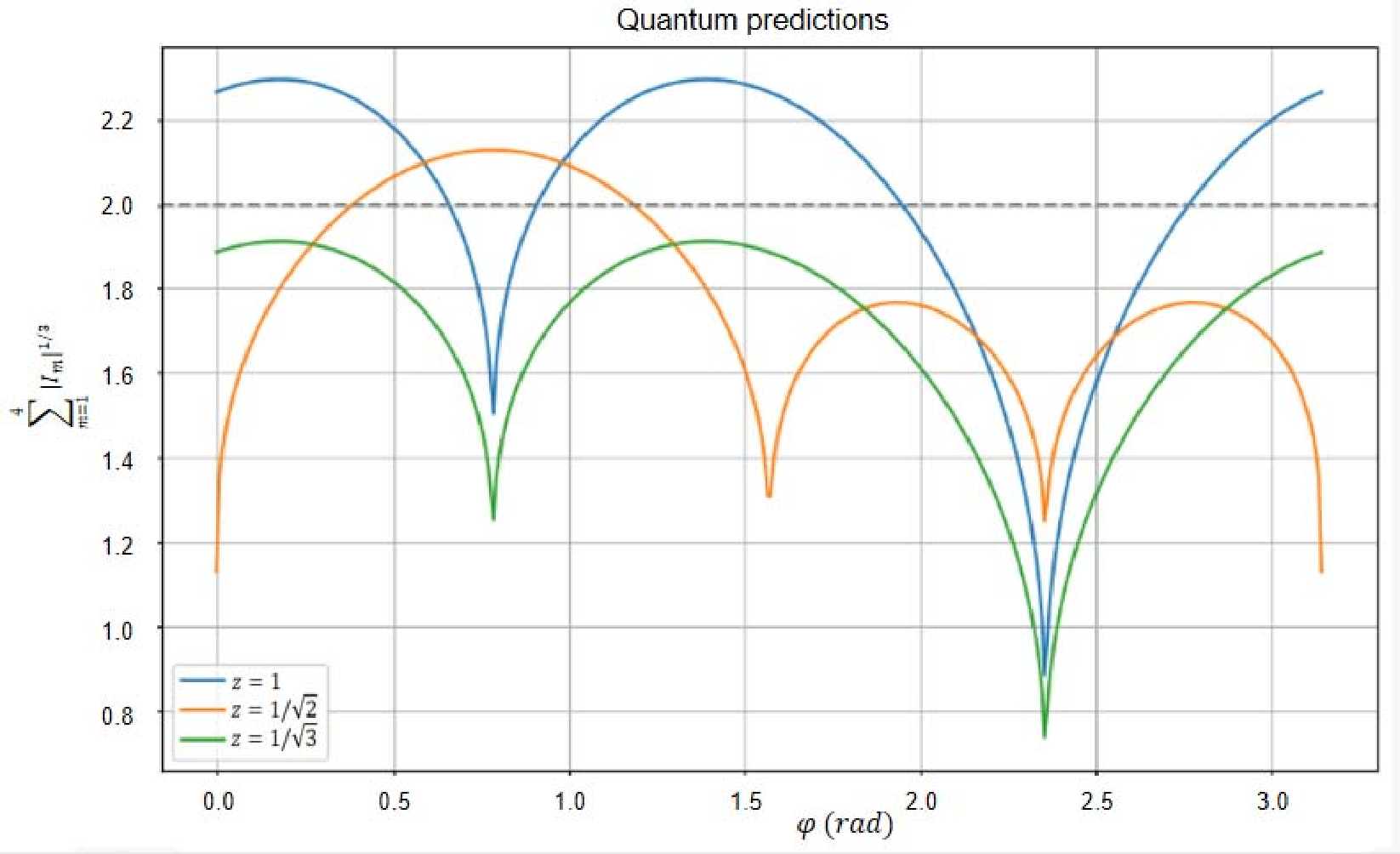}
      \caption{(color online). A plot of the quantum predictions in the range $\varphi \in [0,\pi]$, for $z=1$ (blue line), $z=1/\sqrt{2}$ (orange line), and $z=1/\sqrt{3}$ (green line), respectively.}
      \label{I-m.eps}
\end{figure}

Now, according to Eqs. (\ref{b123}) and (\ref{Psi-000-00}), we calculate the inner products $\langle \widetilde{\Psi}_{b_{1}b_{2}b_{3}}| A^{m} |\widetilde{\Psi}_{b_{1}b_{2}b_{3}}\rangle$. A detailed calculation can be found in the appendix C.
Substituting the results into (\ref{I-m}) leads to
\begin{eqnarray}\label{}
I_{1} = \frac{1}{8} z\sin (2\gamma) \cos [2(\varphi-\varphi_{z})]\sin (\varphi+\frac{\pi}{4}),
\end{eqnarray}
\begin{eqnarray}\label{}
I_{2} = \frac{1}{4} z\sin (2\gamma)\sin (\varphi+\frac{\pi}{4}),
\end{eqnarray}
\begin{eqnarray}\label{}
I_{3} = \frac{1}{4\sqrt{2}} z(1+\sin \theta)\cos (\varphi-\varphi_{z}+\frac{\pi}{4}),
\end{eqnarray}
and
\begin{eqnarray}\label{}
I_{4} = \frac{1}{4\sqrt{2}} z(1+\sin \theta)\sin (\varphi-\varphi_{z}+\frac{\pi}{4}).
\end{eqnarray}
Based on these trilocal correlation quantities, we thus have
\begin{eqnarray}\label{}
\sum_{m=1}^{4} |I_{m}|^{1/3} &=& \frac{\sqrt[3]{|z|}}{2}
[|\sin (2\gamma) \cos [2(\varphi-\varphi_{z})]\sin (\varphi+\frac{\pi}{4})|^{\frac{1}{3}}
+|2\sin (2\gamma)\sin (\varphi+\frac{\pi}{4})|^{\frac{1}{3}} \nonumber \\ &&
+|\sqrt{2} (1+\sin \theta)\cos (\varphi-\varphi_{z}+\frac{\pi}{4})|^{\frac{1}{3}}
+|\sqrt{2} (1+\sin \theta)\sin (\varphi-\varphi_{z}+\frac{\pi}{4})|^{\frac{1}{3}}].
\end{eqnarray}
Using numerical analysis we obtain the maximum value of the quantum prediction $2.2968$ with $\gamma = \pi/4$, $\theta = \pi/2$, $z=1$ (thus $\varphi_{z}=\pi/2$) and $\varphi=0.1781$ rad (or $1.3921$ rad).
Taking $\gamma = \pi/4$ and $\theta = \pi/2$, i.e., $\tau = 1$, we provide a plot of the quantum predictions in the range $\varphi \in [0,\pi]$, for $z=1/\sqrt{3}$, $z=1/\sqrt{2}$, and $z=1$, respectively, as shown schematically in Fig.\ref{I-m.eps}.
It indicates that the quantum predictions vary continuously with $\varphi$, and the trilocal inequality (\ref{trilocal-inequality}) can be violated for $z=1$ and $z=1/\sqrt{2}$.

\section{$N$-qubit symmetric joint measurements}

In this section, we provide the generalized $n$-qubit symmetric joint measurements for $n > 3$.
Consider the single-qubit bases $\{|\pm m_{i}\rangle\}$, $i=0,1,2,3$, two-qubit EJM bases $\{|\Phi_{i}\rangle \}_{i=0}^{3}$ and $\{|\Phi'_{i}\rangle \}_{i=0}^{3}$, as described earlier.
We define the $n$-qubit symmetric joint measurements:
for $n$ is even,
\begin{eqnarray} \label{}
|\Psi_{ij_{1}\cdots j_{k}}\rangle &=&
  \cos \gamma |\Phi_{i}\rangle|\Phi_{j_{1}}\rangle \cdots |\Phi_{j_{k}}\rangle
+ (-1)^{[\frac{i}{2}]} \sin \gamma |\Phi'_{i}\rangle|\Phi'_{j_{1}}\rangle \cdots |\Phi'_{j_{k}}\rangle,
\end{eqnarray}
where $i, j_{1}, \cdots, j_{k} =0,1,2,3$, $k=n/2 -1$;
for $n$ is odd,
\begin{eqnarray} \label{}
|\Psi_{ij_{1}\cdots j_{k}}^{0}\rangle &=&
  \cos \gamma |\Phi_{i}\rangle|\Phi_{j_{1}}\rangle \cdots |\Phi_{j_{k}}\rangle |m_{i}\rangle
+ (-1)^{[\frac{i}{2}]} \sin \gamma |\Phi'_{i}\rangle |\Phi'_{j_{1}}\rangle \cdots |\Phi'_{j_{k}}\rangle|-m_{i}\rangle
\end{eqnarray}
and
\begin{eqnarray} \label{}
|\Psi_{ij_{1}\cdots j_{k}}^{1}\rangle &=&
  \cos \gamma |\Phi_{i}\rangle|\Phi_{j_{1}}\rangle \cdots |\Phi_{j_{k}}\rangle |-m_{i}\rangle
- (-1)^{[\frac{i}{2}]} \sin \gamma |\Phi'_{i}\rangle |\Phi'_{j_{1}}\rangle \cdots |\Phi'_{j_{k}}\rangle|m_{i}\rangle,
\end{eqnarray}
where $i, j_{1}, \cdots, j_{k} =0,1,2,3$, $k=(n-1)/2 -1$.

Note that there are two fundamental components that form the $n$-qubit symmetric joint measurements. For an even $n$, it is depicted with paired two-qubit EJM bases; for an odd $n$, four sets of additional single-qubit base $\{|\pm m_{i}\rangle\}$, $i=0,1,2,3$, are added.
So, a calculation similar to three-qubit EJM basis yields the orthogonality relations
\begin{eqnarray} \label{}
\langle \Psi_{ij_{1}\cdots j_{k}}^{}|\Psi_{i'j'_{1}\cdots j'_{k}}^{}\rangle = \delta_{ii'}\delta_{j_{1}j'_{1}}\cdots \delta_{j_{k}j'_{k}}, ~ i,i',j_{1},j'_{1}, \cdots,j_{k},j'_{k}=0,1,2,3,
\end{eqnarray}
for $n$ is even, and
\begin{eqnarray} \label{}
\langle \Psi_{ij_{1}\cdots j_{k}}^{l}|\Psi_{i'j'_{1}\cdots j'_{k}}^{l'}\rangle = \delta_{ii'}\delta_{j_{1}j'_{1}}\cdots \delta_{j_{k}j'_{k}}\delta_{ll'}, ~ i,i',j_{1},j'_{1}, \cdots,j_{k},j'_{k}=0,1,2,3, ~ l,l'=0,1,
\end{eqnarray}
for $n$ is odd.
Using $\sum_{j_{k}=0}^{3} |\Phi_{j_{k}}\rangle \langle \Phi'_{j_{k}} | = \sum_{j_{k}=0}^{3} |\Phi'_{j_{k}}\rangle \langle \Phi_{j_{k}}|$, in the same way, we have the completeness relations
\begin{eqnarray} \label{}
\sum_{i,j_{1},\cdots,j_{k}=0}^{3}|\Psi_{ij_{1}\cdots j_{k}}^{}\rangle
\langle \Psi_{ij_{1}\cdots j_{k}}^{}| = I,
\end{eqnarray}
for $n$ is even, and
\begin{eqnarray} \label{}
\sum_{i,j_{1},\cdots,j_{k}=0}^{3}(|\Psi_{ij_{1}\cdots j_{k}}^{0}\rangle\langle \Psi_{ij_{1}\cdots j_{k}}^{0}|
                                 +|\Psi_{ij_{1}\cdots j_{k}}^{1}\rangle\langle \Psi_{ij_{1}\cdots j_{k}}^{1}|) = I,
\end{eqnarray}
for $n$ is odd.

\begin{figure}[ht]
      \centering
      \includegraphics[width=4in]{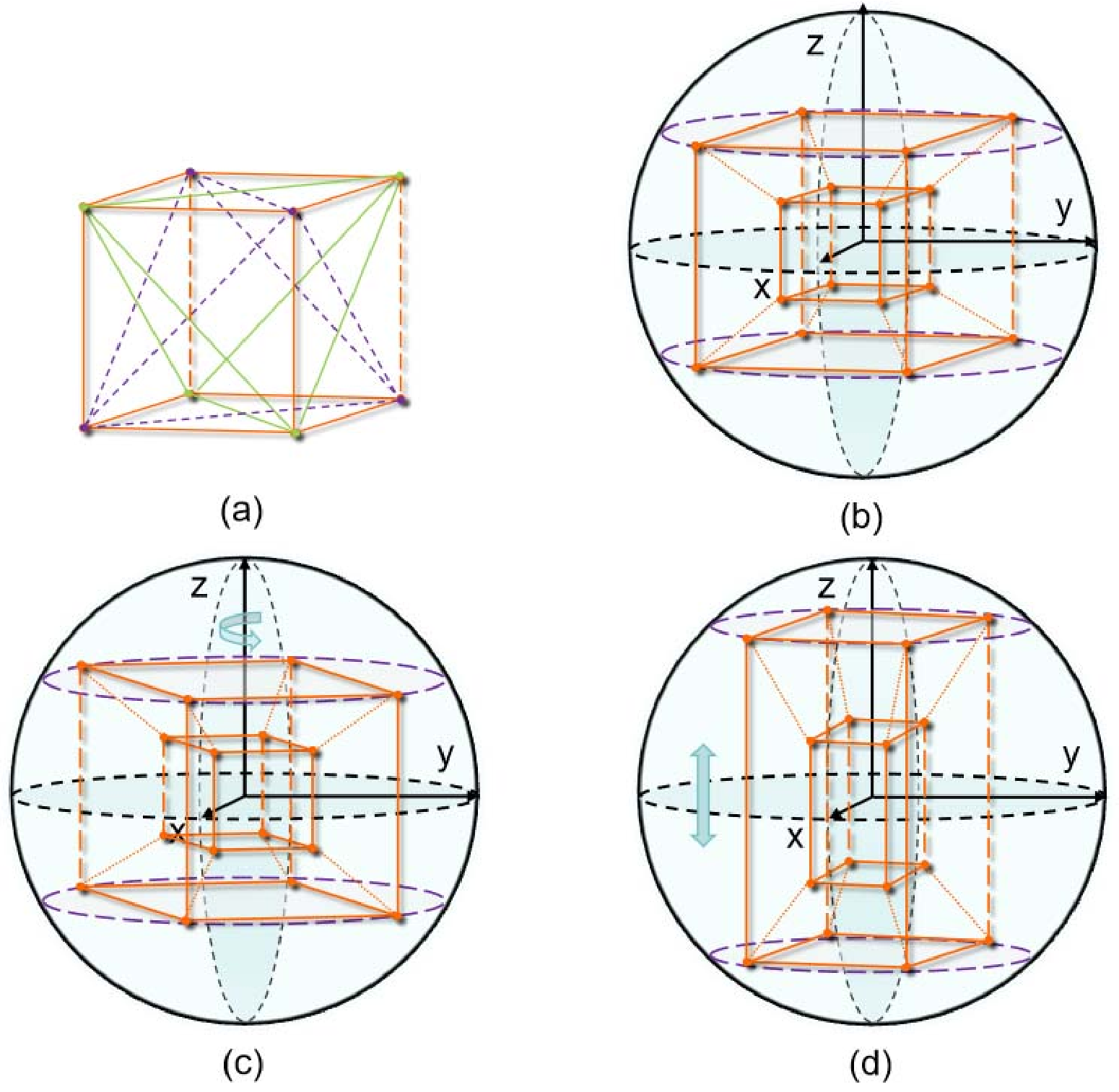}
      \caption{The schematic diagram illustrating hexahedral symmetry of the generalized multi-qubit EJM: (a) The correspondence relationship between the vertices of a hexahedron and two tetrahedra; (b) the reduced states and the corresponding unit vectors to which they point; (c) the set of reduced states associated with an arbitrary rotation around the $z$-axis; and (d) the set of reduced states associated with a stretch along the $z$-axis.}
      \label{symad.eps}
\end{figure}

For an odd-qubit EJM basis, there is one qubit whose reduced states correspond to $\pm \cos(2\gamma) \vec{m}_{i}$, while all the remaining single-qubit reduced states correspond to $[(1/2)\sqrt{1+2\cos^{2} (2\gamma)}\cos \theta] \vec{m'}_{i}$ or $- [(1/2)\sqrt{1+2\cos^{2} (2\gamma)}\cos \theta] \vec{m'}_{i}$.
All single-qubit reductions of even-qubit basis states match the points $[(1/2)\sqrt{1+2\cos^{2} (2\gamma)}\cos \theta] \vec{m'}_{i}$ or $- [(1/2)\sqrt{1+2\cos^{2} (2\gamma)}\cos \theta] \vec{m'}_{i}$. As described earlier in the three-qubit EJM basis, these reductions can form vertices of pairs of mirror-image tetrahedra within the
Bloch sphere, and thus the equations
\begin{eqnarray} \label{}
\sum_{P} P [\sum_{i,j_{1},\cdots, j_{k},l}\langle \Psi_{ij_{1}\cdots j_{k}}^{l}|\vec{\sigma} \otimes I^{\otimes (n-1)}|\Psi_{ij_{1}\cdots j_{k}}^{l} \rangle] = 0
\end{eqnarray}
hold for all odd-qubit EJM bases, and
\begin{eqnarray} \label{}
\sum_{P} P [\sum_{i,j_{1},\cdots, j_{k}}\langle \Psi_{ij_{1}\cdots j_{k}}^{}|\vec{\sigma} \otimes I^{\otimes (n-1)}|\Psi_{ij_{1}\cdots j_{k}}^{} \rangle] = 0
\end{eqnarray}
hold for all even-qubit EJM bases, where $P$ runs over all permutations of the position of ``$\vec{\sigma}$.''
Moreover, in fact, the vertices of these mirror-image tetrahedra can always form a rectangular parallelepiped.
Based on this, it can also be interpreted as rectangular-parallelepiped (or hexahedral) symmetry.
For the sake of clarity, we are providing a diagram here to illustrate the hexahedral symmetry for the current $n$-qubit EJM, as shown in Fig.\ref{symad.eps}.

\section{Conclusion}

In summary, we studied a generalization of the two-qubit EJM \cite{Gisin-EJM2019,TGB-EJM2021,3-parameter-EJM2025} to one involving $n$ qubits.
To do this, we first defined a three-qubit symmetric joint measurement basis by using the three-parameter two-qubit EJM basis and four sets of single-qubit bases. Since these basis states correspond to the vertices of a tetrahedron located on the Bloch sphere \cite{3-parameter-EJM2025}, the current architecture ensures that it has the tetrahedral (or hexahedral) symmetry for the single-qubit reductions. In addition to symmetry, we also demonstrated its application to a trilocal star-shaped network.
More importantly, as a general method, it can be naturally extended to $n$-qubit EJM, preserving the elegant symmetry.

In applying it to the star-network nonlocality, however, we found that the example provided does not currently result in the maximum violation \cite{PhysRevA.90.062109(2014)}.
Note that, in the given measurement settings, the quantum prediction, in general, varies with the parameters $\gamma$, $\theta$, $z$, and $\varphi$.
By numerically optimizing the measurement settings, it may be possible to further enhance the quantum prediction value, potentially bringing it to or near its maximum value of $2\sqrt{2}$.
Nevertheless, it is indeed an interesting issue of principle to develop the symmetric quantum joint measurements on multiple qubits, which could potentially lead to new insights into the fundamental nature of quantum measurement and advance the field of quantum information processing.

\begin{acknowledgements}
This work was supported by
the Hebei Science and Technology Program Foundation of China under Grant No. 246Z0902G,
the National Natural Science Foundation of China under Grant No. 62271189,
and the Hebei High-level Talents Foundation of China under Grant No: A202101002.
\end{acknowledgements}

\appendix
\section{Proof of the orthogonality and completeness for the three-qubit EJM}

We first see the orthogonality of the three-qubit EJM basis states.
Note that each single-qubit basis state $|m^{}_{i}\rangle$ is orthogonal to $|-m^{}_{i}\rangle$
and two-qubit basis states satisfy $\langle \Phi_{i}|\Phi_{j}\rangle = \delta_{ij}, ~i,j=0,1,2,3$. Also, for $i=0,1$, $|\Phi'_{i}\rangle = |\Phi_{i+2}\rangle$, and for $i=2,3$, $|\Phi'_{i}\rangle = |\Phi_{i-2}\rangle$.
Thus, for $i=j$, we can directly see that
\begin{eqnarray} \label{}
\langle\Psi_{i}^{k}|\Psi_{i}^{l}\rangle =  \cos^{2} \gamma \langle \Phi_{i}|\Phi_{i}\rangle \delta_{kl} +
\sin^{2} \gamma \langle \Phi'_{i}|\Phi'_{i}\rangle \delta_{kl} = \delta_{kl}, ~ i=0,1,2,3,k,l=0,1.
\end{eqnarray}

For $i\neq j$, it is divided into three cases: (i) $i=0,2$, and $j=1,3$, or vice versa; (ii) $i,j=0,2$; and (iii) $i,j=1,3$.
For $i=0,2$ and $j=1,3$ (or vice versa), using $\langle \Phi_{i}|\Phi_{j}\rangle = 0$, we can easily verify that $\langle\Psi_{i}^{k}|\Psi_{j}^{l}\rangle = 0$.
For $i,j=0,2$, we calculate $\langle\Psi_{0}^{k}|\Psi_{2}^{l}\rangle, k,l=0,1,$ as
\begin{eqnarray} \label{}
\langle\Psi_{0}^{0}|\Psi_{2}^{1}\rangle = -\langle\Psi_{0}^{1}|\Psi_{2}^{0}\rangle
= \sin\gamma \cos\gamma (\langle m_{0}|m_{2}\rangle + \langle -m_{0}|-m_{2}\rangle),
\end{eqnarray}
and
\begin{eqnarray} \label{}
\langle\Psi_{0}^{k}|\Psi_{2}^{k}\rangle =  \sin\gamma \cos\gamma (\langle -m_{0}|m_{2}\rangle -\langle m_{0}|-m_{2}\rangle), ~ k=0,1.
\end{eqnarray}
Note that $\langle -m_{0}|-m_{2}\rangle = -\langle m_{0}|m_{2}\rangle = z \text{i}$, $\langle -m_{0}|m_{2}\rangle = \langle m_{0}|-m_{2}\rangle = -\text{i}\sqrt{1-z^{2}}$. So we have $\langle\Psi_{0}^{k}|\Psi_{2}^{l}\rangle = 0$, and thus  $\langle\Psi_{2}^{l}|\Psi_{0}^{k}\rangle = (\langle\Psi_{0}^{k}|\Psi_{2}^{l}\rangle)^{\dagger} =0$.
Similarly, for $i,j=1,3$, using $\langle -m_{1}|-m_{3}\rangle = -\langle m_{1}|m_{3}\rangle  = z \text{i}$ and  $\langle -m_{1}|m_{3}\rangle = \langle m_{1}|- m_{3}\rangle = \text{i}\sqrt{1-z^{2}}$, one can obtain $\langle\Psi_{1}^{k}|\Psi_{3}^{l}\rangle = \langle\Psi_{3}^{k}|\Psi_{1}^{l}\rangle = 0$, $k,l=0,1$.

Thus we have the orthogonality relations
\begin{eqnarray} \label{}
\langle\Psi_{i}^{k}|\Psi_{j}^{l}\rangle = \delta_{ij}\delta_{kl}, ~~i,j=0,1,2,3, k,l=0,1.
\end{eqnarray}

Now let us show the completeness relation by calculating the outer product  $\sum_{i,k}|\Psi_{i}^{k}\rangle \langle \Psi_{i}^{k}|$, $i=0,1,2,3$; $k=0,1$.

For $i=0,1$, a straightforward calculation by Eqs. (\ref{Psi-i-0}) and (\ref{Psi-i-1}) shows that
\begin{eqnarray} \label{}
\sum_{k=0,1}|\Psi_{i}^{k}\rangle \langle \Psi_{i}^{k}|
&=& (\cos^{2}\gamma |\Phi_{i}\rangle \langle \Phi_{i}|
   +\sin^{2}\gamma |\Phi_{i+2}\rangle \langle \Phi_{i+2}|)
   \otimes (|m_{i}\rangle \langle m_{i}|+|-m_{i}\rangle \langle -m_{i}|)
\\ \nonumber  &&
+\cos\gamma \sin\gamma [(|\Phi_{i}\rangle \langle \Phi_{i+2}|
                        -|\Phi_{i+2}\rangle \langle \Phi_{i}|)
                \otimes (|m_{i}\rangle \langle -m_{i}| - |-m_{i}\rangle \langle m_{i}|)].
\end{eqnarray}
Similarly, for $i=2,3$, we have
\begin{eqnarray} \label{}
\sum_{k=0,1}|\Psi_{i}^{k}\rangle \langle \Psi_{i}^{k}| &=& (\cos^{2}\gamma |\Phi_{i}\rangle \langle \Phi_{i}| +\sin^{2}\gamma |\Phi_{i-2}\rangle \langle \Phi_{i-2}|)
\otimes (|m_{i}\rangle \langle m_{i}|+|-m_{i}\rangle \langle -m_{i}|)
\\ \nonumber  &&
-\cos\gamma \sin\gamma
       [(|\Phi_{i}\rangle \langle \Phi_{i-2}|-|\Phi_{i-2}\rangle \langle \Phi_{i}|)
\otimes (|m_{i}\rangle \langle -m_{i}| - |-m_{i}\rangle \langle m_{i}|)],
\end{eqnarray}
Using completeness relation $|m_{i}\rangle \langle m_{i}|+|-m_{i}\rangle \langle -m_{i}|=I$ and summing over the ensemble yields
\begin{eqnarray} \label{}
\sum_{i=0}^{3}\sum_{k=0}^{1}|\Psi_{i}^{k}\rangle \langle \Psi_{i}^{k}| &=&
(\sum_{i=0}^{3} |\Phi_{i}\rangle \langle \Phi_{i}|) \otimes I
+ \cos\gamma \sin\gamma (|\Phi_{0}\rangle \langle \Phi_{2}|-|\Phi_{2}\rangle \langle \Phi_{0}|)
\otimes \sum_{i=0,2} (|m_{i}\rangle \langle -m_{i}| - |-m_{i}\rangle \langle m_{i}|)
\\ \nonumber  &&
+ \cos\gamma \sin\gamma (|\Phi_{1}\rangle \langle \Phi_{3}|-|\Phi_{3}\rangle \langle \Phi_{1}|)
\otimes \sum_{i=1,3} (|m_{i}\rangle \langle -m_{i}| - |-m_{i}\rangle \langle m_{i}|).
\end{eqnarray}
We note that
$|m_{i}\rangle \langle -m_{i}| - |-m_{i}\rangle \langle m_{i}| = \text{e}^{\text{i}\varphi_{i}}|1\rangle \langle 0| - \text{e}^{-\text{i}\varphi_{i}}|0\rangle \langle 1|$
and thus
$\sum_{i=0,2}(|m_{i}\rangle \langle -m_{i}| - |-m_{i}\rangle \langle m_{i}|) =
 \sum_{i=1,3}(|m_{i}\rangle \langle -m_{i}| - |-m_{i}\rangle \langle m_{i}|) = 0$.
So, combining these results yields
\begin{eqnarray} \label{}
\sum_{i=0}^{3}\sum_{k=0}^{1}|\Psi_{i}^{k}\rangle \langle \Psi_{i}^{k}|  =
 \sum_{i=0}^{3}|\Phi_{i}\rangle \langle \Phi_{i}| \otimes I = I.
\end{eqnarray}

\section{Reductions of the three-qubit EJM basis states}

We here calculate the reductions of the three-qubit EJM basis states by tracing out two qubits.
Basing directly on the Eqs. (\ref{Psi-i-0}) and (\ref{Psi-i-1}) and noting that $\langle m_{i}|I|m_{i}\rangle = \langle -m_{i}|I|-m_{i}\rangle = 1$, $\langle m_{i}|I|-m_{i}\rangle = \langle -m_{i}|I|m_{i}\rangle = 0$, we have
\begin{eqnarray} \label{}
\langle \Psi_{i}^{k}|\vec{\sigma} \otimes I \otimes I |\Psi_{i}^{k}\rangle &=&
 \cos^{2}\gamma \langle \Phi_{i}|\vec{\sigma} \otimes I |\Phi_{i}\rangle
+\sin^{2}\gamma \langle \Phi_{i+2}|\vec{\sigma} \otimes I |\Phi_{i+2}\rangle, ~ i=0,1,
\end{eqnarray}
and
\begin{eqnarray} \label{}
\langle \Psi_{i}^{k}|\vec{\sigma} \otimes I \otimes I |\Psi_{i}^{k}\rangle &=&
 \cos^{2}\gamma \langle \Phi_{i}|\vec{\sigma} \otimes I |\Phi_{i}\rangle
+\sin^{2}\gamma \langle \Phi_{i-2}|\vec{\sigma} \otimes I |\Phi_{i-2}\rangle, ~ i=2,3.
\end{eqnarray}

Note that
\begin{eqnarray} \label{}
\langle \Phi_{i}|\vec{\sigma} \otimes I |\Phi_{i}\rangle &=&
\frac{1}{\sqrt{2}} \cos \theta
(\cos (\varphi_{i}-\varphi_{z}), \sin (\varphi_{i}-\varphi_{z}), (-1)^{i}/\sqrt{2}), ~ i=0,1,2,3,
\end{eqnarray}
$\cos (\varphi_{i\pm2}-\varphi_{z}) = -\cos (\varphi_{i}-\varphi_{z})$ and $\sin (\varphi_{i\pm2}-\varphi_{z}) = -\sin (\varphi_{i}-\varphi_{z})$.
Substituting we obtain
\begin{eqnarray} \label{}
\langle \Psi_{i}^{k}|\vec{\sigma} \otimes I \otimes I |\Psi_{i}^{k}\rangle &=&
\frac{1}{2} \cos \theta (\sqrt{2}\cos (2\gamma)\cos(\varphi_{i}-\varphi_{z}), \sqrt{2}\cos (2\gamma)\sin(\varphi_{i}-\varphi_{z}), (-1)^{i}), i=0,1,2,3.
\end{eqnarray}
Let \begin{eqnarray} \label{}
\vec{m'}_{i}=\frac{1}{\sqrt{1+2\cos^{2} (2\gamma)}} (\sqrt{2}\cos (2\gamma)\cos(\varphi_{i}-\varphi_{z}), \sqrt{2}\cos (2\gamma)\sin(\varphi_{i}-\varphi_{z}),(-1)^{i}),
\end{eqnarray}
and then we have
\begin{eqnarray} \label{}
\langle \Psi_{i}^{k}|\vec{\sigma} \otimes I \otimes I |\Psi_{i}^{k}\rangle =
\frac{1}{2}[\sqrt{1+2\cos^{2} (2\gamma)}\cos \theta]  \vec{m'}_{i}.
\end{eqnarray}
A similar calculation for $\langle \Psi_{i}^{k}|I \otimes \vec{\sigma} \otimes I|\Psi_{i}^{k}\rangle$ yields
\begin{eqnarray} \label{}
\langle \Psi_{i}^{k}|I \otimes \vec{\sigma} \otimes I|\Psi_{i}^{k}\rangle =
- \frac{1}{2}[\sqrt{1+2\cos^{2} (2\gamma)}\cos \theta] \vec{m'}_{i}.
\end{eqnarray}

To proceed, by using $\langle \Phi_{i}|\Phi_{j}\rangle = \delta_{ij}$, we have
\begin{eqnarray} \label{}
\langle \Psi_{i}^{0}| I \otimes I \otimes \vec{\sigma}|\Psi_{i}^{0}\rangle &=&
 \cos^{2}\gamma \langle  m_{i}|\vec{\sigma} | m_{i}\rangle
+\sin^{2}\gamma \langle -m_{i}|\vec{\sigma} |-m_{i}\rangle,
\end{eqnarray}
and
\begin{eqnarray} \label{}
\langle \Psi_{i}^{1}| I \otimes I \otimes \vec{\sigma}|\Psi_{i}^{1}\rangle &=&
 \cos^{2}\gamma \langle  -m_{i}|\vec{\sigma} | -m_{i}\rangle
+\sin^{2}\gamma \langle m_{i}|\vec{\sigma} |m_{i}\rangle,
\end{eqnarray}
for $i=0,1,2,3$. Then using $\langle  m_{i}|\vec{\sigma}| m_{i}\rangle =  \vec{m}_{i}$ and $\langle - m_{i}|\vec{\sigma}|- m_{i}\rangle = - \vec{m}_{i}$,  we have
\begin{eqnarray} \label{}
\langle \Psi_{i}^{k}|I \otimes I\otimes \vec{\sigma}|\Psi_{i}^{k}\rangle
= (-1)^{k}\cos (2\gamma) \vec{m}_{i},
\end{eqnarray}
where $\vec{m}_{i}=(\sqrt{1-z_{i}^{2}}\cos \varphi_{i},\sqrt{1-z_{i}^{2}}\sin \varphi_{i},z_{i}), ~ i=0,1,2,3$.

\section{Derivation of the trilocal correlation quantities}

By using one-to-one correspondences (\ref{Psi-000-00}), we here rewrite the three-qubit EJM basis as
\begin{eqnarray} \label{Psi0}
|\Psi_{000}\rangle =
\cos \gamma |\Phi_{0}\rangle|m_{0}\rangle + \sin \gamma |\Phi_{2}\rangle|-m_{0}\rangle, ~~~
|\Psi_{001}\rangle =
\cos \gamma |\Phi_{0}\rangle|-m_{0}\rangle - \sin \gamma |\Phi_{2}\rangle|m_{0}\rangle,
\end{eqnarray}
\begin{eqnarray} \label{Psi1}
|\Psi_{010}\rangle =
\cos \gamma |\Phi_{1}\rangle|m_{1}\rangle + \sin \gamma |\Phi_{3}\rangle|-m_{1}\rangle, ~~~
|\Psi_{011}\rangle =
\cos \gamma |\Phi_{1}\rangle|-m_{1}\rangle - \sin \gamma |\Phi_{3}\rangle|m_{1}\rangle,
\end{eqnarray}
\begin{eqnarray} \label{Psi2}
|\Psi_{100}\rangle =
\cos \gamma |\Phi_{2}\rangle|m_{2}\rangle - \sin \gamma |\Phi_{0}\rangle|-m_{2}\rangle, ~~~
|\Psi_{101}\rangle =
\cos \gamma |\Phi_{2}\rangle|-m_{2}\rangle + \sin \gamma |\Phi_{0}\rangle|m_{2}\rangle,
\end{eqnarray}
\begin{eqnarray} \label{Psi3}
|\Psi_{110}\rangle =
\cos \gamma |\Phi_{3}\rangle|m_{3}\rangle - \sin \gamma |\Phi_{1}\rangle|-m_{3}\rangle, ~~~
|\Psi_{111}\rangle =
\cos \gamma |\Phi_{3}\rangle|-m_{3}\rangle + \sin \gamma |\Phi_{1}\rangle|m_{3}\rangle.
\end{eqnarray}
Next, we take $\langle \widetilde{\Psi}_{000}| A^{1} |\widetilde{\Psi}_{000}\rangle$ as an example to give a detailed derivation to calculate the trilocal correlation quantities.

We calculate $|\widetilde{\Psi}_{000}\rangle$ and have
\begin{eqnarray} \label{}
|\widetilde{\Psi}_{000}\rangle &=& \frac{1+\text{i}\sqrt{3z^{2}-1}}{2\sqrt{3z^{2}}}
[-d^{+}_{1}\text{e}^{-\text{i}(\frac{3}{2}\varphi-\varphi_{z})}|000\rangle
-d^{-}_{0}\text{e}^{-\text{i}(\frac{1}{2}\varphi-\varphi_{z})}|001\rangle
-d^{-}_{1}(r_{-}^{\theta})^{\ast}\text{e}^{-\text{i}\frac{\varphi}{2}}|010\rangle
-d^{+}_{0}(r_{-}^{\theta})^{\ast}\text{e}^{\text{i}\frac{\varphi}{2}}|011\rangle
\nonumber \\ &&
-d^{-}_{1}(r_{+}^{\theta})^{\ast}\text{e}^{-\text{i}\frac{\varphi}{2}}|100\rangle
-d^{+}_{0}(r_{+}^{\theta})^{\ast}\text{e}^{\text{i}\frac{\varphi}{2}}|101\rangle
+d^{+}_{1}\text{e}^{\text{i}(\frac{1}{2}\varphi-\varphi_{z})}|110\rangle
+d^{-}_{0}\text{e}^{\text{i}(\frac{3}{2}\varphi-\varphi_{z})}|111\rangle],
\end{eqnarray}
where $d^{\pm}_{0} = (\cos \gamma \sqrt{1+z} \pm \sin\gamma \sqrt{1-z})/\sqrt{2}$, $d^{\pm}_{1} = (\cos \gamma \sqrt{1-z} \pm \sin\gamma \sqrt{1+z})/\sqrt{2}$,
and $(r_{\pm }^{\theta})^{\ast} =(1\pm \text{e}^{-\text{i}\theta})/\sqrt{2}$.
Substituting $A^{1} = \sigma_{x}\sigma_{x}\sigma_{x}$ yields
\begin{eqnarray} \label{}
\langle \widetilde{\Psi}_{000}| \sigma_{x}\sigma_{x}\sigma_{x} |\widetilde{\Psi}_{000}\rangle &=& -\frac{1}{2} [\sqrt{1-z^{2}}\cos (2\gamma) + z \sin (2\gamma)] \cos [ 2(\varphi-\varphi_{z})]  \cos \varphi.
\end{eqnarray}

Similarly, we calculate all of the inner products $\langle \widetilde{\Psi}_{b_{1}b_{2}b_{3}}| \sigma_{x}\sigma_{x}\sigma_{x} |\widetilde{\Psi}_{b_{1}b_{2}b_{3}}\rangle$.
Also, considering definition of bits $b_{1},b_{2},b_{3}$  (\ref{b123}) we have
\begin{eqnarray} \label{}
M^{1}_{0}= |\Psi_{001}\rangle \langle \Psi_{001}| + |\Psi_{010}\rangle \langle \Psi_{010}|
+ |\Psi_{101}\rangle \langle \Psi_{101}| + |\Psi_{110}\rangle \langle \Psi_{110}|
\end{eqnarray}
and
\begin{eqnarray} \label{}
M^{1}_{1}= |\Psi_{000}\rangle \langle \Psi_{000}| + |\Psi_{011}\rangle \langle \Psi_{011}|
+ |\Psi_{100}\rangle \langle \Psi_{100}| + |\Psi_{111}\rangle \langle \Psi_{111}|.
\end{eqnarray}
So, substituting these into (\ref{I-m}) we have
\begin{eqnarray}\label{}
I_{1} = \frac{1}{8} z\sin (2\gamma) \cos [2(\varphi-\varphi_{z})]\sin (\varphi+\frac{\pi}{4}).
\end{eqnarray}
A calculation similar to $I_{1}$ yields
\begin{eqnarray}\label{}
I_{2} = \frac{1}{4} z\sin (2\gamma)\sin (\varphi+\frac{\pi}{4}),
\end{eqnarray}
\begin{eqnarray}\label{}
I_{3} = \frac{1}{4\sqrt{2}} z(1+\sin \theta)\cos (\varphi-\varphi_{z}+\frac{\pi}{4}),
\end{eqnarray}
and
\begin{eqnarray}\label{}
I_{4} = \frac{1}{4\sqrt{2}} z(1+\sin \theta)\sin (\varphi-\varphi_{z}+\frac{\pi}{4}).
\end{eqnarray}

\end{document}